\documentclass[sigchi]{acmart}

\AtBeginDocument{%
  \providecommand\BibTeX{{%
    \normalfont B\kern-0.5em{\scshape i\kern-0.25em b}\kern-0.8em\TeX}}}

\setcopyright{none}
\renewcommand\footnotetextcopyrightpermission[1]{} 

\setcopyright{acmcopyright}
\copyrightyear{2022}
\acmYear{2022}
\acmDOI{10.1145/1122445.1122456}

\acmConference{}
\acmBooktitle{}
\acmPrice{15.00}
\acmISBN{978-1-4503-XXXX-X/18/06}

\usepackage{caption}
\usepackage{subcaption}
\begin{document}

\newcommand{\revise}[1]{\textcolor{blue}{#1}}
\newcommand\mengyu[1]{\textcolor{blue}}
\newcommand\zhenchang[1]{\textcolor{red}{\textbf{Zhenchang}:{#1}}}
\newcommand\chunyang[1]{\textcolor{violet}{\textbf{Sherry}:#1}}
\newcommand\jieshan[1]{\textcolor{orange}{\textbf{Jieshan}:#1}}

\title{Enhancing Virtual Assistant Intelligence: Precise Area Targeting for Instance-level User Intents beyond Metadata}


\author{Mengyu Chen}
\email{mengyu.chen@data61.csiro.au}
\affiliation{%
  \institution{Data61, CSIRO}
  \country{Australia}
  }

\author{Zhenchang Xing}
\email{zhenchang.xing@data61.csiro.au}
\affiliation{%
  \institution{Data61, CSIRO \& Australian National University}
  \country{Australia}
  }

\author{Jieshan Chen}
\email{jieshan.chen@data61.csiro.au}
\affiliation{%
  \institution{Data61, CSIRO}
  \country{Australia}
  }

\author{Chunyang Chen}
\email{chunyang.chen@monash.edu}
\affiliation{%
  \institution{Faculty of Information Technology, Monash University}
  \country{Australia}
  }

\author{Qinghua Lu}
\email{qinghua.lu@data61.csiro.au}
\affiliation{%
  \institution{Data61, CSIRO}
  \country{Australia}
  }


\begin{abstract}
Virtual assistants have been widely used by mobile phone users in recent years. Although their capabilities of processing user intents have been developed rapidly, virtual assistants in most platforms are only capable of handling pre-defined high-level tasks supported by extra manual efforts of developers. However, instance-level user intents containing more detailed objectives with complex practical situations, are yet rarely studied so far. In this paper, we explore virtual assistants capable of processing instance-level user intents based on pixels of application screens, without the requirements of extra extensions on the application side. We propose a novel cross-modal deep learning pipeline, which understands the input vocal or textual instance-level user intents, predicts the targeting operational area, and detects the absolute button area on screens without any metadata of applications. We conducted a user study with 10 participants to collect a testing dataset with instance-level user intents. The testing dataset is then utilized to evaluate the performance of our model, which demonstrates that our model is promising with the achievement of 64.43\% accuracy on our testing dataset.
\end{abstract}





\maketitle
\begin{figure*}
    \centering
    \includegraphics[width=\textwidth]{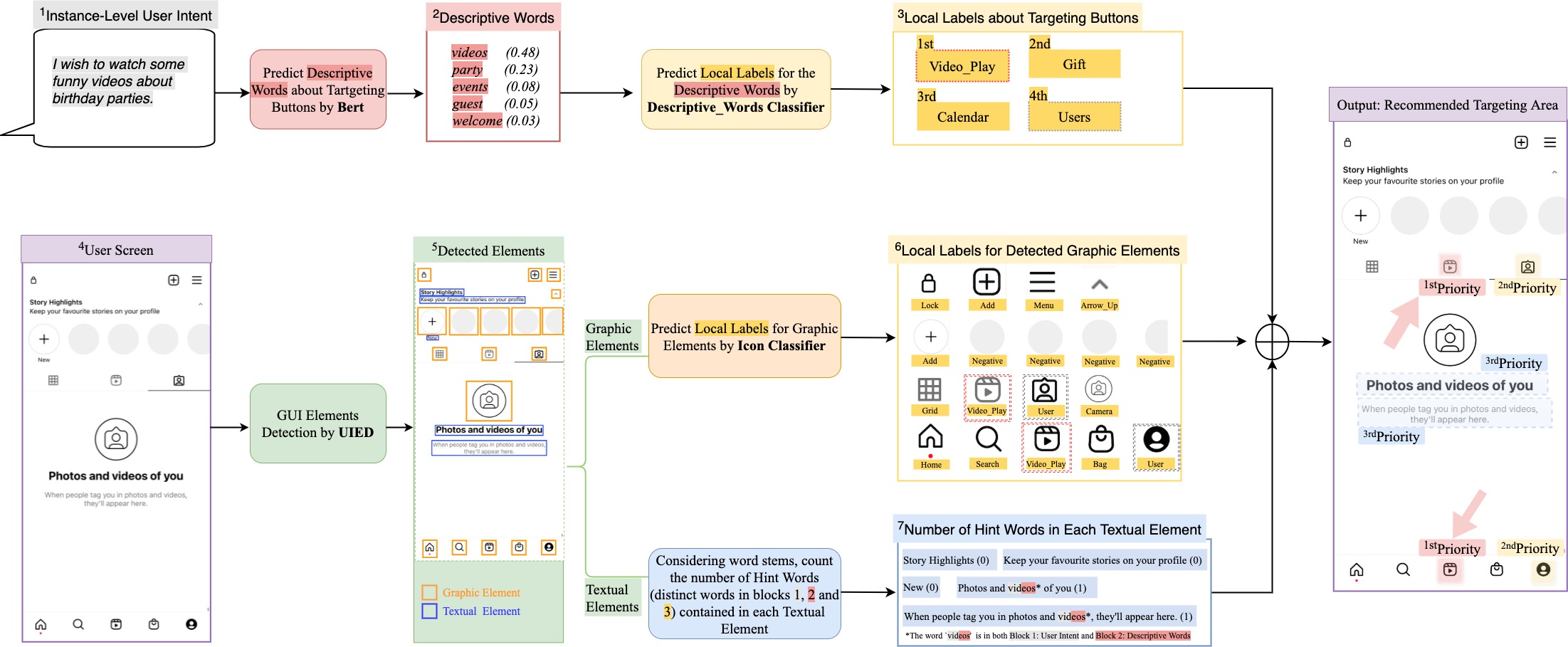}
    \caption{The flowchart of the model. The flow displayed on the upper side demonstrates the process for user intent: the oral or textual instance-level intent (colored in gray) is firstly processed by Bert, to predict the descriptive words with corresponding possibilities for the Targeting Buttons (colored in red); the descriptive words are then classified with local labels by Descriptive Word Classifier for targeting buttons, and the printed local labels are sorted in descending priorities orders (colored in yellow); distinct words in the block with tag 1, 2 and 3 are regarded as Hint Word, that would be used in the lower side flow. The flow displayed on the lower side demonstrates the process for the oriented user screen: GUI elements including graphical elements and textual elements are detected by UIED (colored in green); the graphical elements detected by UIED are classified with local labels by Icon Classifier (colored in orange); the Hint Word are searched and counted among each textual element detected by UIED in word stem manner (colored in blue). Combining the textual and graphical results from the upper and lower flows together, the targeting operational areas are detected. Details of the working flow are in Section~\ref{sec_approach}.}
    \label{fig_flow_chart}
\end{figure*}
\section{Introduction}
Virtual assistants, such as Siri, have become an integral part of our daily lives since their inception in 2011 \cite{wikipedia_2022}. These intelligent systems have transformed the way we interact with our devices, allowing us to perform a variety of tasks through simple voice commands. For instance, setting an alarm, checking the weather, or making a phone call can all be accomplished without any physical interaction with the device. However, the capabilities of these virtual assistants are primarily confined to built-in applications, limiting the range of tasks they can perform.

While platform-specific mechanisms like Siri Shortcuts~\cite{apple_support_2020} and Android App Actions~\cite{android_developers} have been introduced to extend the functionality of virtual assistants, these solutions heavily rely on additional development efforts~\cite{uied}. Consequently, if a user's intent falls outside the predefined service scope, the virtual assistant is unable to fulfill the request and instead, resorts to displaying related websites or admitting its limitations. This gap between the growing diversity of user intents and the restricted functionalities of virtual assistants is a significant challenge in the current app market.

Another critical issue is the understanding and detection of user intents. Virtual assistants often struggle to comprehend and process complex user intents, particularly those at the instance-level, which include detailed scenarios. Given the myriad ways in which humans can express a purpose, it is virtually impossible to predefine a corpus to cover all possible user intents.

These aforementioned limitations can hinder the full potential of human-computer interaction, particularly when it comes to interacting with third-party applications installed on the device. 
Therefore, to enhance the interaction between virtual assistants and installed apps, it is crucial to develop a system that can identify the precise operational area on the screen based on user intents.

Previous research has focused on enhancing the capabilities of virtual assistants~\cite{r_10, r_24, r_28, r_46, r_49, r_58}. A common approach is to grant the assistant access to the metadata of installed apps. However, this method requires additional efforts from developers and may lead to privacy and security issues. Moreover, the performance of machine learning-based models that map voice commands to actions is dependent on a large amount of labeled data, which is labor-intensive to collect and label.

To address these challenges, we propose a novel pipeline that leverages natural language processing and computer vision techniques to precisely target the operational area for instance-level user intents beyond metadata. Our model is unsupervised, eliminating the need for predefined user intents and intent-button label data. It also does not require access to metadata, ensuring user privacy. Furthermore, it can be universally adopted for installed apps without requiring additional efforts from developers or end-users.

In this paper, we present a novel pipeline that combines natural language processing and computer vision techniques to precisely target the operational area for instance-level user intents beyond metadata. Our model is unsupervised, eliminating the need for predefined user intents and intent-button label data. It also does not require access to metadata, ensuring user privacy. Furthermore, it can be universally adopted for installed apps without requiring additional efforts from developers or end-users.

We evaluate our model using a user study to collect a diverse, representative, and complete testing dataset. The dataset includes various user intents and the ground truth of the operational area from a range of ranked app categories on the App Store for iPhone, iPad, and Mac. Our model achieves an accuracy of $64.43\%$ on the collected testing UIs and $58.56\%$ on the UIs with positive Mosaic UIs.

Our contributions are as follows:
\begin{itemize}
    \item We propose a novel task to empower the virtual assistant to interact with installed apps, with the ability to precisely target operational areas for instance-level user intents without access to metadata.
    \item To the best of our knowledge, we are the first to adopt the Bert MLM model for thoroughly understanding user intents and accurately detecting user objectives, while precisely predicting descriptive words for targeting areas.
    \item We designed a novel pipeline through unsupervised learning combining natural language processing techniques and computer vision techniques, without built-in intents and intent-button training dataset required.
    \item We collected a dataset with diversity, representativeness, and completeness, containing instance-level user intents and precise targeting operational areas on iOS UIs for testing automation.
\end{itemize}

\section{Related Work}
This section presents an overview of virtual assistants, their limitations, and the design goals for our proposed approach.

\subsection{Virtual Assistants}

Virtual assistants such as Siri for iOS and Google Assistant for Android have become ubiquitous across a plethora of smart devices. These assistants facilitate hands-free and even eyes-free interactions, automating everyday tasks through simple voice commands from users~\cite{wulf2014hands, r_58, uist}. However, the services offered by these virtual assistants are primarily based on built-in apps, thereby limiting their functional scope. The integration of installed apps with virtual assistants necessitates additional manual efforts from developers. Despite these efforts, there remain significant limitations on both Android and iOS platforms~\cite{uist}. 

\subsection{Limitations of Virtual Assistants}

Google Assistant supports only a limited number of built-in intents~\cite{uist}. Developers can use App Actions~\cite{android_developers} to enable installed apps to interact with Google Assistant, requiring the addition of an extra file (actions.xml) to declare the specific built-in intents that can be utilized. These built-in intents are pre-defined in the Android system and cannot be expanded by app developers. Consequently, Google Assistant cannot automatically process custom tasks that fall outside the scope of built-in intents. Although the list of built-in intents can be expanded by Android system developers, it would necessitate considerable manual efforts to investigate the custom capabilities of apps in the rapidly evolving app market and update the corresponding system documents. 

Siri on iOS interacts with installed apps in a manner similar to Google Assistant, requiring an Intent Definition File that declares the app's capabilities (i.e., iOS intents)~\cite{uist}. Siri also allows developers to encode custom intents, necessitating a mechanism that donates the shortcut to Siri every time the user intent is supported ~\cite{apple_developer_documentation, uist}. Developers can also link app functionality to Siri via suggested Shortcuts~\cite{apple_support_2020}, which requires manual configuration from end-users.

Despite the considerable manual efforts from developers and end-users, the added intent file or edited shortcut is app-specific and cannot be automatically adopted by other apps with similar functionalities. Furthermore, the exposure of app capabilities is platform-specific. For the same app serving on different platforms like iOS and Android, developers must make efforts to interact with different virtual assistants respectively. 

Moreover, to the best of our knowledge, virtual assistants only have app-level access to installed apps, which treats a certain app as a tool to complete a specific task for end-users. In contrast, many everyday tasks are on UI-level, which heavily relies on certain UI contents. The UI-level user intents thus cannot be processed by current virtual assistants.

\subsection{Design Goals of our Approach}

In light of the limitations of current virtual assistants, we propose several design goals for our virtual assistant in this paper: 1) zero-effort is required from both developers and end-users to expose app capabilities to the virtual assistant; 2) the virtual assistant is cross-app and cross-platform; 3) the virtual assistant is capable of processing custom tasks and UI-level tasks. In addition to addressing the aforementioned limitations, we also set extra design goals to further enhance the capabilities of virtual assistants: 4) thoroughly and accurately understand instance-level user intents; 5) precisely identify the target operation area of user intent; 6) operate based on UIs only without any metadata of apps to protect the privacy and security of end-users.
\section{Approach}
\label{sec_approach}
This section provides a comprehensive overview of our model architecture.

\subsection{Model Architecture}

Our model takes a screenshot and an instance-level vocal or textual user intent as input. This intent clarifies the user's objective in detail. We have developed an innovative deep-learning pipeline model to address this problem.

Initially, our model comprehends and identifies the intent from natural language. It then predicts the most probable words for the target button names in general language, along with their corresponding probabilities. This is achieved using the Masked Language Model of Bert~\cite{devlin2018bert}. The words predicted by Bert are then mapped to local labels from a local 80-class label list through a word-to-label translation agent. This agent also calculates the search priorities for the recommended local labels using multiple voting algorithms. The output from the word-to-label translation agent comprises the target labels to be searched on the screen.

Simultaneously, all icons and texts on the input screens are detected and captured using a GUI elements detection toolkit, UIED~\cite{uied}. The captured icons are then classified into a set of local labels from the aforementioned local 80-class label list through an icon classifier. The target labels recommended by the word-to-label translation agent are then searched among the set of labels predicted by the icon classifier in descending order of their priorities.

If none of the target labels are found among the predicted labels from icons, the text contents on the input screen detected by UIED are considered. Each distinct word in the textual user intent is recognized as a token for text searching. The words predicted by Bert, as well as the target local labels recommended by the word-to-label translation agent, are also considered as tokens. These identified tokens are then searched over the text contents detected on the screen. The text area containing the largest number of tokens is identified as the target operating area for the user intent.

\subsection{User Intent Detection and Prediction}

Our objective is to identify the target operational button on an input screen based on the user intent. Unlike object detection tasks in the computer vision field, where the input user intent is a defined 'label', our task requires the model to interpret the natural language description of user intent to determine the user's goal and provide hints for the target area.

Unlike high-level user intents, instance-level user intents contain more detailed practical scenarios. For example, intents such as \textit{`send money'}, \textit{`set an alarm'}, and \textit{`listen to music'} are high-level, while \textit{`I wish to check what I have added into the trolley'}, \textit{`I am interested in fashion and would like to dive deep into that category'}, and \textit{`I am trying to find what translation records I have saved for future use'} are instance-level. Furthermore, user intent can be expressed in a multitude of ways. For instance, consider the objective of wanting to speak during an online meeting. This could be articulated as \textit{`I want to speak'}, \textit{`I wish to unmute myself'}, or \textit{`I would like to use my microphone'}, among other variations. The flexibility of natural language expressions makes the set of user intents infinite, which cannot be processed directly as a classification problem like most high-level intent tasks. Instead, our model should be able to detect the intent from the instance-level description and produce hints or predictions for the target operational area. For example, the goal of speaking during an online meeting. In our experience with mobile applications, the target operational area is usually linked to 'voice' functions. These clues about the target operational area are instrumental in directing attention towards pertinent icons among the multitude of components present on the screen.

A proficient natural language processor with extensive experience in mobile applications is highly desirable. However, no current model can predict an operational target area based on user intent. Our approach draws inspiration from the training process of Bert (Bidirectional Encoder Representations from Transformers) \cite{devlin2018bert}. Bert employs a Masked Language Model (MLM) that randomly masks 15\% of all WordPiece tokens in each input sentence sequence, predicting the masked vocabulary id based solely on context. This demonstrates Bert's proficiency in predicting a masked word within a sentence. For instance, given a sentence with a special \textbf{`[MASK]'} mark, the MLM of a pre-trained Bert treats the \textit{`[MASK]'} as a missing vocabulary and can predict a list of words with corresponding probabilities. Furthermore, Bert's extensive training on a vast amount of textual materials equips it with the `knowledge' to provide suggestions for numerous queries it has previously `learned'. For the sentence \textit{`The capital of the UK is [MASK].'}, the top five vocabularies predicted by Bert (`bert-base-uncased' version) are \textit{`london'} (probability 0.265), \textit{`manchester'} (probability 0.094), \textit{`edinburgh'} (probability 0.070), \textit{`birmingham'} (probability 0.060), and \textit{`bristol'} (probability 0.034) \cite{devlin2018bert}.

We have incorporated the `bert-base-uncased' model into our system, which is pre-trained on a dataset called BookCorpus, comprising 11,038 unpublished books and English Wikipedia (excluding lists, tables, and headers) \cite{devlin2018bert}. This model is case-insensitive, meaning it does not distinguish between \textit{`english'} and \textit{`English'}. We selected it because its training dataset is sufficiently capable of learning an inner representation of the English language \cite{devlin2018bert}. To equip the Bert model with expertise on mobile application guidelines, we fine-tuned it using 50 mobile application handbooks downloaded from official mobile application websites. Leveraging Bert's prediction ability, we employed it in a novel way to provide hints for the target operational button by appending the text — \textbf{`, so the [MASK] icon should be clicked.'} to the original user intent. For instance, if the original user input is \textit{`I wish to check what I have added into the trolley'}, the input for the Bert model would be \textit{`I wish to check what I have added into the trolley, so the [MASK] icon should be clicked.'}. Consequently, Bert outputs a list of candidates for the \textit{`[MASK]'} unit with corresponding probabilities in descending order. We take the top five vocabularies predicted by Bert as the input for the next step. Please note that our model can process English user intent only. All vocal user intents are first transformed to text by \textbf{`speech-to-text'} technology.

\subsection{Element Detection and Classification on Screen}

With the vocabulary predicted by Bert, the logical next step is to identify screen components related to the provided descriptive words. However, unlike humans who instinctively distinguish different items in an image, computers perceive images as mere combinations of pixel values, devoid of any sense of the objects displayed. To find the target button, our model must be capable of identifying various components on a screen.

We employed User Interface Element Detection (UIED)~\cite{uied} to detect elements on the input screen. UIED is a toolkit specifically designed for GUI element detection. Utilizing UIED allows us to detect the locations of both graphic and textual content on a screen.

For the detected graphic elements, we trained a CNN icon classifier to assign a label to each icon based on its pattern. The training dataset was derived from icon images on the Common Mobile/Web App Icons' from the Kaggle website \cite{test.ai_2018}, which contains 106 classes, with 153380 images. The Kaggle dataset also includes a `negative' class with 10564 images that do not belong to any of the other 105 classes. Since the Kaggle dataset is designed for both web and mobile Apps, some categories are unsuitable for our tasks. For instance, icons in the cursor' category rarely appear on mobile screens, so we removed these categories from the training dataset. Moreover, as stated in the official description of the Common Mobile/Web App Icons' dataset, many icons are misclassified. For example, a large number of arrow right', arrow up', and arrow down' images are incorrectly categorized as arrow left', which significantly limits the classifier's capability and performance. We manually cleaned the images in each category and moved the images with incorrect labels into the correct category. Considering that the detected graphic elements may include images not intended for clicking (e.g., background images, profile pictures, images inserted into textual content, or illustrative diagrams), we retained the negative' category to distinguish these image contents from normal icons. After manual data cleaning, the number of images used for the classifier is 26970 in 80 categories, with 79 icon classes and 1 \textit{`negative'} class.

We trained an 8-layer CNN model with 1000 hidden neurons as the icon classifier, with a momentum of 0.5 and kernel size of 5, with a batch size of 64 for 30 epochs. This model achieved the best performance compared to models with other experimental settings, given the constraints of our hardware device (i9-10900k CPU and GeForce RTX 3090 GPU with 24G memory). It reached 95.57\% accuracy on the testing dataset. With the icon classifier, the graphic elements detected by UIED are classified into one of 80 categories. Consequently, each graphic element resembling a mobile icon is predicted with one of the labels from the 79 icon classes based on its pattern, while the ones similar to images in the content are labeled as `negative'.

For the textual elements on the screen, both their locations and textual contents are captured by UIED, which are also utilized for identifying the target operational icon in the subsequent steps.

\section{Cross-Modality Bridging}
Within the purview of this study, natural language predictions generated by BERT serve to suggest potential identifications of the target operational button. Simultaneously, UIED's screen detection capabilities manifest an array of plausible candidates for the suggested button, which encompasses both graphical and textual components. Moving forward, we introduce a pioneering cross-modal alignment mechanism, with the aim of pinpointing the element most closely harmonizing with BERT's descriptive word output. This mechanism will specifically navigate through the graphic and textual elements detected by UIED independently, thereby ensuring a comprehensive search.

\subsection{An Infeasible Naive Search in Visual Elements}

We want to find the visual elements which align best with hints from Bert. Note that, the descriptive words provided by Bert pertain solely to the functional attributes of the requisite buttons, rather than their design patterns. 

With the aforementioned icon classifier, each graphic element detected by UIED is predicted with one label among the $80$ local classes. We ignored the elements with the \textit{`negative'} label, since they are not regarded as normal buttons by the icon classifier. To find the target button, a naive approach is to measure the similarity between the label of each detected graphic element and the descriptive words produced by Bert. However, this naive approach highly depends on the linguistic contents of labels, which may result in significant bias. To be specific, the assigned labels of graphic elements originate from category names within the training dataset. The category names are employed as distinguishing markers for images across distinct classes, which are not compulsorily related to the content of images inside each category. This is evidenced by the frequent use of numerals as labels in many classification tasks. 

Although labels could be manually replaced with words pertaining to the content of images within each category, it is improbable to find a word that accurately and unequivocally describes the functions of images within each category. As an example, the natural words \textit{`cart', `shop', `buy'}, and \textit{`purchase'} may all denote a category of icons associated with shopping. In the training dataset, we selected \textit{`cart'} as the label for shopping-related icons and `house' as the label for home-related icons. In the user intent, \textit{`I wish to purchase a new umbrella'}, one predicted descriptive word from BERT is \textit{`Store'}. On the user screen, there is an icon bearing a \textit{`shopping cart'} pattern with the actual functionality - \textit{purchasing}, and an icon bearing a \textit{`house'} pattern with the actual functionality - \textit{homepage}. Then
the \textit{`house'} icon might be erroneously selected as the target button, since compared with the word \textit{`cart'}, the word \textit{`house'} has higher similarity with the descriptive word \textit{`Store'}.

To mitigate this issue, we developed a Descriptive-Words Classifier to match the predictive words suggested by BERT with one or more labels from the local 80 labels, entirely irrespective of label linguistics. Here, the local 80 labels are exactly the same as the labels produced by the icon classifier.

\subsection{An Preliminary Approach for Visual Elements Searching}

Our initial approach of the Descriptive-Words Classifier is that, given a natural descriptive word of a category of icons, e.g. `alarm', a certain amount of images with the keyword \textit{`alarm icon'} are automatically downloaded  through online search engines. Then the Descriptive-Words Classifier is utilized to predict a label for each downloaded image. For each user intent of the model, multiple descriptive words are predicted by BERT, as mentioned in the previous section. For each word predicted by BERT, the Descriptive-Words Classifier is utilized to predict labels for images downloaded with this keyword. As a result, a list of labels as predicted by the classifier. These predicted labels, arranged in descending order of frequency, serve as the output labels for BERT's descriptive words.

Although effective for many user intents, the principal issue with this approach is the excessive runtime per user intent—over $20$ seconds—owing to speed limitations imposed by online search engines, making it unsuitable for practical applications. To enhance the speed and accuracy, we revised the approach by replacing the online search engine with a local dataset complemented by multiple voting agents for label selection.

The advanced Descriptive-Words Classifier integrates a local dataset with 13 voting agents, each of which votes for one or more labels based on its unique voting algorithm. The labels, arranged in descending order based on their respective votes, are presented as the output. The detailed structure of the Descriptive-Words Classifier is as follows.

\begin{figure*}
    \centering
    \includegraphics[width=0.8\textwidth]{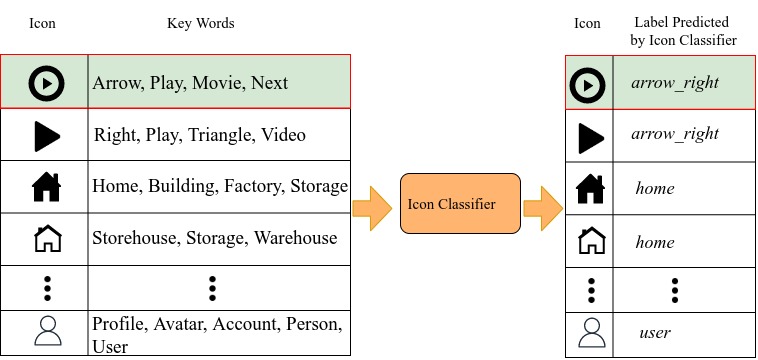}
    \caption{An illustrative example of the source of data in the generated local database.}
    \label{fig_noun_project}
\end{figure*}

\subsection{Local Database Generation}

The local dataset was created using data from the Noun Project, which encompasses a comprehensive array of information about icons, such as names, designers, searching words for related icons, and so forth. Herein, the searching words as the keywords for finding corresponding icons \cite{noun_project}. As illustrated in the left-hand table in Figure \ref{fig_noun_project}, each image is associated with multiple search words or keywords. For instance, the icon in the first row is linked to four keywords: \textit{`Arrow'}, \textit{`Play'}, \textit{`Movie'}, and \textit{`Next'}. While the icon in the second row is with keywords \textit{`Right'}, \textit{`Play'}, \textit{`Triangle'}, and \textit{`Video'}. It is important to note that icons with similar patterns may have distinct keywords, as demonstrated by the icons in the third and fourth rows.

\begin{figure}[h]
    \centering
    \includegraphics[width=0.45\textwidth]{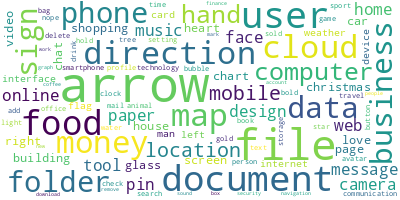}
    \caption{The Word Cloud presenting the top 100 descriptive words with the highest frequencies in the local database.}
    \label{fig_cloud_word}
\end{figure}

\begin{figure*}
    \centering
    \includegraphics[width=\textwidth]{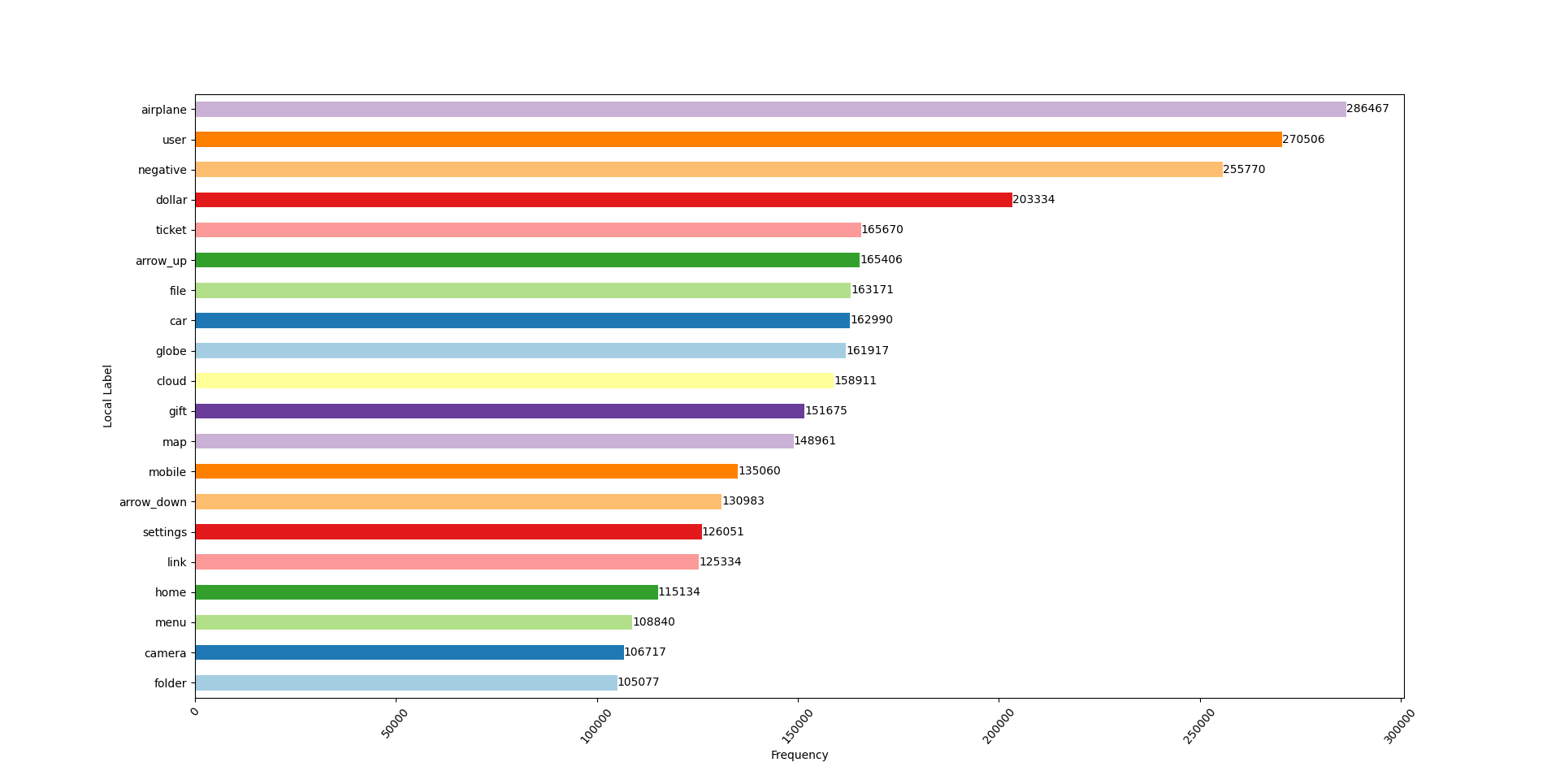}
    \caption{The top 20 labels with the most descriptive words in the local dataset. }
    \label{fig_bar_label}
\end{figure*}

In order to populate the local database, the previously discussed icon classifier was harnessed to predict a label for each icon image. Consequently, each icon image within the Noun Project database was assigned a local label from a pool of 80 classes. For instance, the icon delineated in the first row of the table was allocated the local label \textit{`arrow\_right'}. Viewing the icon's image as an intermediary, we accordingly generated four word-label pairings, specifically \textit{(`Arrow', `arrow\_right')}, $(\textit{`Play'}, \textit{`arrow\_right'})$, (\textit{`Movie'}, \textit{`arrow\_right'}), and (\textit{`Next'}, \textit{`arrow\_right'}). These derived word-label pairs were integrated into the local database for analytical evaluation. We incorporated a total of $1,547,847$ images from the Noun Project, which collectively provided $6,348,857$ descriptive words. Of these, $214,377$ words were unique, culminating in $6,348,857$ word-label pairs stored within the local dataset. Figure \ref{fig_cloud_word} illustrates a word cloud comprising the top $100$ descriptive words from the database, ranked by frequency - the larger the font size, the higher the frequency. Figure \ref{fig_bar_label} depicts the top $20$ labels with the highest associated descriptive word count within the database.

\subsection{Multiple Voting Agents}

Given the predictive words and their corresponding probabilities as predicted by Bert, our objective is to utilize the local database to identify a local label that aligns most closely with these predictive words. Several strategies can be employed to match the local labels, each focusing on distinct features and potentially yielding different local labels. Take, for instance, the predictive words \textit{`call'}, \textit{`phone'}, \textit{`user'}, \textit{`contact'}, and \textit{`talk'}, each with their associated probabilities. One strategy might involve identifying all word-label pairs within the database that match any of these five predictive words, and selecting the local label with the highest frequency as the final choice. Alternatively, another strategy might consider each of the five words separately, selecting the local label that most frequently pairs with each predictive word in the database. For example, the local labels selected for \textit{`mobile'}, \textit{`phone'}, \textit{`user'}, \textit{`contact'}, and \textit{`talk'} might be \textit{`call'}, \textit{`call'}, \textit{`user'}, \textit{`file'}, and \textit{`volume'}, respectively. In this case, the local label 'call' appears twice, which is more than any other label, and is therefore selected as the final local label. Both strategies are feasible, but may result in differing final local labels. It is also worth noting that both strategies are largely influenced by the composition of the local database's data source, meaning changes to the source that introduce more pairs of certain descriptive words or local labels could significantly impact the outcome of each strategy. 

In an effort to enhance generalization and select the local label independently of data source composition, we designed a group of voting agents that consider various features. This includes eight deterministic algorithms and five non-deterministic algorithms, which collectively improve fairness and reduce biases caused by the data source. The specifics of each voting algorithm are as follows.

\subsubsection{Deterministic Voting Algorithms}

In Approach 1.1, 1.2, and 1.3, for each word predicted by Bert, we identify a candidate local label. This candidate is chosen based on its highest frequency pairing with the respective word in the database. For instance, given the user intent, \textit{`If click the [MASK] icon, it adds the item to the shopping trolley'}, we receive the following predicted descriptive words in Table~\ref{table:word_prob}.
\begin{table}[h]
\centering
\begin{tabular}{|c|c|c|}
\hline
\textbf{Index $(k)$} & {Word$(w_k)$} & \textbf{Probability$(p_k)$} \\
\hline
$1$ & $item$ & $0.217$ \\
\hline
$2$ & $store$ & $0.0759$ \\
\hline
$3$ & $shopping$ & $0.0497$ \\
\hline
$4$ & $product$ & $0.0415$ \\
\hline
$5$ & $purchase$ & $0.0374$ \\
\hline
\end{tabular}
\caption{An example of descriptive words predicted by Bert with corresponding probabilities.}
\label{table:word_prob}
\end{table}

When we input $w_1 = 'item'$ as a descriptive word into the database, we can find $3,058$ word-label pairs with 'items', among which the word-label pair with the highest frequencies are (\textit{`item'}, \textit{`airplane'}), with $128$ appearances among the $3,058$ pairs, we thus get the percentage $q_1$ of $`(w_1, l_1)'$ pair among all $w_1$ related pairs as $q_1 = \frac{128}{3058} = 4.186\%$. In the same way, we can get the related labels and percentages for $w_2, w_3, w_4$ and $w_5$ respectively:

 \begin{itemize}
 \item $l_1 = 'airplane'$ with $q_1 = 16.882\%$
\item $l_2 = 'cart'$ with $q_2 = 16.882\%$
\item $l_3 = 'cart'$ with $q_3 = 26.298\%$
\item $l_4 = 'barcode'$ with $q_4 = 7.617\%$ 
\item $l_5 = 'cart'$ with $q_5 = 22.665\%$.
\end{itemize}

So far, we have got $5$ local label candidates together with their percentages.
\\\\
\textbf{Agent 1.1}: Only focus on the 5 label candidates from $l_1$ to $l_5$, vote for the one with the maximum appearances among the $5$ labels. 

\textit{In this case, there are 3 distinct labels, `airplane’, `cart’ and `barcode’, with the number of appearances $1$, $3$ and $1$ respectively. So the label with the maximum appearance is `\textbf{cart}', which gets one vote from Approach 1.1.}
\\\\
\textbf{Agent 1.2}: For each distinct label $l'$ from $l_1$ to $l_5$, find its accumulated percentage $f(l') = \sum_{ l_k = l'}q_k$ and vote for the one with the maximum $f(l')$ value.

\textit{In this case, there are 3 unique labels, `airplane’, `cart’ and `barcode’, with
\begin{itemize}
    \item $f('airplane') = q_1 =4.186\% $
    \item $f('cart') = q_2 + q_3+ q_5 =16.882\% + 26.298\% +22.665\% =65.845\%$
    \item $f('barcode') = q_4 =7.617\%$
\end{itemize}
So the label with the maximum  $f(l')$ value is `\textbf{cart}', which gains one vote from Approach 1.2.}
\\\\
\textbf{Agent 1.3}: For each distinct label $l'$ from $l_1$ to $l_5$, find the accumulated value of its probability multiplies its percentage $g(l') = \sum_{l_k = l'}p_k q_k$, where probability is provided by Bert. Vote for the candidate label with the maximum $g(l')$ value.

\textit{We have, 
\begin{itemize}
    \item $g('airplane') = p_1q_1 = 0.908\%$
    \item $g('cart') = p_2q_2 +p_3q_3+ p_5q_5 =3.436\% $
    \item  $g('barcode') = p_4q_4 =0.316\%$
\end{itemize}
So the label with the maximum $g(l')$ value is ‘\textbf{cart}’, which gains one vote from Approach 1.3.}
\\\\
In subsequent strategies, there is no requirement to identify a corresponding local label candidate for each word predicted by Bert. 
\\\\

\textbf{Agent 1.4}: For each word $w_k$, find all labels $l_k^m$ paired with $w_k$, together with the count $c_k^m$ of the label $l_k^m$ stored in the data set. Select the label $l'$ with the maximum count $c'$.

\textit{In this case, the $5$ words are \textit{`item’}, \textit{`store’}, \textit{`shopping’}, \textit{`product’} and \textit{`purchase’}. Among word-label pairs containing any one of these $5$ words, the frequencies of local labels in non-increasing order are: $3,911$ for \textit{`cart'}, $2,117$ for \textit{`bag'}, $932$ for \textit{`dollar'}, ..., where \textit{`cart'}, \textit{`bag'} and \textit{`dollar'} are local labels. The label \textbf{`cart'} with the maximum number $3,911$ of images thus gains one vote from Approach 1.4.}
\\\\

 \textbf{Agent 1.5}: For each word $w_k$ provided by Bert, find all labels $l_k^m$ paired with $w_k$, together with the percentage $q_k^m$ that label $l_k^m$ stored in the dataset. The label $l'$ with the maximum percentage $q'$ gains one vote from Approach 1.5.

\textit{In this case, the label \textbf{`cart'} with the maximum sum of percentage $0.7334$ is selected. }
\\\\

\textbf{Agent 1.6}:  For each distinct label $l'$ from $l_1$ to $l_5$, find its averaged accumulated percentage $f(l') = avg(\sum_{l_k = l'}  q_k)$. Vote for the one with the maximum $f(l')$ value.

\textit{In this case, there are $3$ distinct labels, `airplane’, `cart’ and `barcode’, with
\begin{itemize}
    \item $f('airplane') = q_1 =4.186\%$
    \item $f('cart') = \frac{q_2 + q_3+ q_5 }{3} =\frac{16.882\% + 26.298\% +22.665\% }{3} =21.948\%$
    \item $f('barcode') = q_4 =7.617\%$
\end{itemize}
So the label with the maximum  $f(l')$ value is \textbf{`cart'}, which gains one vote from Approach 1.6.}
\\\\

\textbf{Agent 1.7}: For each unique label $l'$ from $l_1$ to $l_5$, find the averaged accumulated value of its probability multiplies its percentage $g(l') = avg(\sum(l_k = l')  p_k\times q_k)$. Vote for the one with the maximum $g(l')$ value.

\textit{In this case, we have
\begin{itemize}
    \item $g('airplane') = p_1q_1 = 0.2174.186\% = 0.908\%$
    \item $g('cart') = \frac{p_2q_2 +p_3q_3+ p_5q_5}{3}=1.145\%$
    \item  $g('barcode')=p_4q_4 =0.04157.617\% =0.316\%$
\end{itemize}
  So the label with the maximum $g(l')$ value is \textbf{`cart'}, which gains one vote from Approach 1.7.}
\\\\

\textbf{Agent 1.8}: For each word $w_k$, find all labels $l_k^m$ together with the percentage $q_k^m$ label $l_k^m$ stored in the dataset. Select the label $l'$ with the maximum averaged accumulated percentage $q'$. 
\\\\
\subsubsection{Random Voting Algorithms.}

The non-deterministic strategies follow a similar process. For each word $w$ from $w_1$, $w_2$, $w_3$, $w_4$, and $w_5$ predicted by Bert, Approach 2.1 randomly selects $10$ word-label pairs that contain $w$, and casts a vote for the label that appears most frequently among these $10$ pairs. Approaches 2.2, 2.3, 2.4, and 2.5 operate analogously, with the sole distinction being the number of randomly selected pairs: $20$, $50$, $100$, and $200$ respectively.
\\\\
\subsubsection{Decision of Multiple Agents}

Each voting algorithm casts a vote for a single local label, with the exception of cases where multiple labels equally fulfill an approach's criterion. Subsequently, the local labels receiving at least one vote are ranked in descending order based on the count of their votes. This ranked list of voted local labels is returned as the output of the Descriptive-Word Classifier, corresponding to the descriptive words provided by Bert.

Take the aforementioned examples, the local labels voted by the 13 agents are summarized as follows.
\begin{itemize}
    \item \textit{`cart'} with $9$ votes by Agent 1.1 to 1.8 and Agent 2.5
    \item \textit{`bag'} with $2$ votes by Agent 2.1 and 2.4
    \item \textit{`dollar'} with $1$ vote by Agent 2.2
    \item  \textit{`barcode'} with $1$ vote by Agent 2.3
\end{itemize}

Consequently, given the descriptive words predicted by Bert along with their corresponding probabilities as presented in Table~\ref{table:word_prob}, our Descriptive-Words Classifier generates the local labels 'cart', 'bag', 'dollar', and 'barcode'. These labels are arranged in descending order according to the frequency of their appearances.

\subsection{Use Output Lables of Descriptive-Words Classifier to Search Visual Elements}

Recall that the Icon Classifier assigns one of the local labels to each graphic element detected by UIED on the screen. To identify the target graphic element, we commence by searching for elements that bear the highest priority predicted label, which, in the given instance, is `cart'. In cases where the highest priority label is absent among the detected graphic elements, the search proceeds to the label with the next highest priority, and so forth, until a graphic element bearing the desired label is identified and deemed as the target operational button. 

If no detected elements carry any label from the Descriptive-Word Classifier's output, the textual elements detected by UIED are put into consideration.    

\subsection{Search in Textual Elements}

From the prior steps, for each user intention, there are three components that offer textual hints about the target area: 1) the original user intention, which is a text sentence transformed from the user's verbal input; 2) descriptive words predicted by the BERT model, and 3) the labels predicted by the aforementioned Descriptive-Word Classifier. Each distinct word across these three sources is considered as a token.

Take, for example, the user intent being \textit{`I plan to add the item to the shopping cart'}. Upon inputting this intention into BERT, the predictive descriptive words obtained are \textit{`item'}, \textit{`store'}, \textit{`shopping'}, \textit{`product'}, and \textit{`purchase'}. Subsequently, these descriptive words are input into the Descriptive-Word Classifier, and the predicted labels are \textit{`cart'}, \textit{`bag'}, \textit{`dollar'}, \textit{`barcode'}.

The distinct words from these three sets of tokens are: \textit{`I'}, \textit{`plan'}, \textit{`to'}, \textit{`add'}, \textit{`the'}, \textit{`item'}, \textit{`shopping'}, \textit{`item'}, \textit{`store'}, \textit{`product'}, \textit{`purchase'}, \textit{`cart'}, \textit{`bag'}, \textit{`dollar'}, \textit{`barcode'}.

The next step involves filtering out non-meaningful tokens such as \textit{`to'} and \textit{`the'}, as they could occur multiple times on the screen and may lead to unsatisfactory suggestions from our model. The remaining words serve as search tokens to locate within each text element detected on the screen. By counting the number of tokens present within each textual element, a priority can be assigned to elements with a higher token count, increasing the likelihood of their being highlighted on the screen.

\section{User Studies}
To the best of our knowledge, there is no public dataset with instance-level user intents suitable for testing our model. To address this, we conducted a user study to gather test data from volunteers experienced in smartphone usage for over five years. To assess our model's ability to handle user intents in real-world scenarios, we simulated user interactions using screenshots from freely available apps across various categories on the App Store (Australia), adhering to specific regulations. Recognizing that users' understanding of on-screen icons and their expression of intents can vary, we included a diverse sample of participants. This included both males and females from five different age groups, residing in five different countries, speaking a total of five different languages, and holding six different positions. During the user studies, volunteers were asked to interpret testing screens and propose user intents based on their understanding of the semantics of the displayed icons. They also rated the congruence between their expectations of selected icon patterns and the corresponding text content. The user intents provided by the participants were collected and annotated, forming a testing dataset to automatically evaluate our model's performance.

\subsection{Diversity of Participants}
The primary objective of our user study is to gather test data from users to assess our model's performance in real-world applications. User intents, as provided by each participant, are influenced not only by their understanding of icon semantics but also by their unique styles of expressing intents in natural language.

Regarding semantic understanding, there is no standard guide teaching users which icons correspond to specific functionalities. Consequently, users develop their understanding of icons primarily through personal experience with app exploration. This suggests that users with diverse app experiences may interpret the same icon differently. Factors such as the user's country of residence significantly impact their comprehension of icon semantics, as app designs and functionalities can vary considerably across different regional App Stores. Other factors, including gender, age, and occupation, also notably influence users' most frequently used apps.

In terms of the natural language styles used to express user intents, language proficiency cannot be overlooked, in addition to the factors mentioned above.

\begin{figure}[h]
    \centering
    \includegraphics[width=0.45\textwidth]{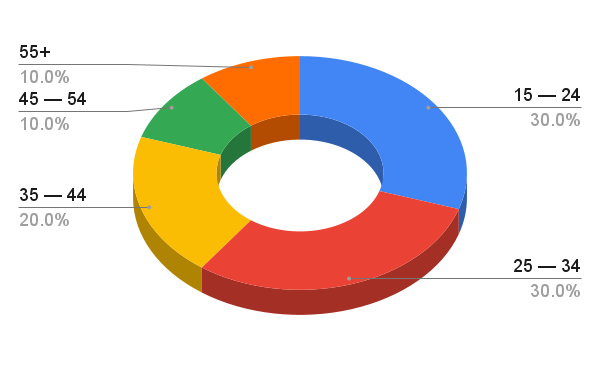}
    \caption{Age distribution of user study participants.}
    \label{fig_pie_user}
\end{figure}

Given these considerations, we invited 10 individuals with over five years of smartphone usage to voluntarily participate in our user studies. These participants currently reside in five different countries: Australia, Canada, China, the United Kingdom, and the United States, listed alphabetically. The group comprises two females and eight males. The age distribution, as shown in Fig.~\ref{fig_pie_user}, ranges from 23 to 56. Participants' occupations include students, algorithm scientists, research fellows, technical sales, and financial analysts. Collectively, the users are proficient in five languages: English, Chinese, Japanese, Korean, and Russian. However, all participants are fluent in English.

\subsection{App Screen Collection}

To assess our model's ability to handle tasks in real-world virtual assistant scenarios, we explored free apps across all ranked categories on the Apple Store for iPhone, iPad, and Mac. Given that our model currently only recognizes user intents in English, we selected a region in the App Store where English is the official language. Although our model can process tasks based on screen icons without relying on text content, characters in languages like Chinese or Korean are often misinterpreted as icon patterns, which hampers icon recognition performance. We plan to develop a more advanced model capable of handling other languages in future studies.

Due to certain restrictions, such as the requirement of American telephone numbers for registration in many apps in the App Store (US), we selected Australia as the App Store region.

We explored all free ranked categories in the App Stores on iPhone, iPad, and Mac. For the iPhone App Store, out of the 28 total categories, we excluded five categories (Apple Watch Apps', AR Apps', Developer Tools', Graphics \& Design', and News') that lack an official ranked list, and the Safari Extension' category, which is not suitable for our tasks. This left us with 22 categories for screen collection. For the iPad App Store, we explored 23 categories, including the `News' category. The Mac App Store, unlike the iPhone and iPad, has 21 categories, all with an official ranked list. In total, we downloaded 198 apps: 66 for iPhone, 69 for iPad, and 63 for Mac.

For each free and ranked category, we downloaded and installed the top-ranked app, the 20th ranked app, and a randomly selected app. Apps for iPhone and iPad were downloaded on February 28, 2022, while Mac apps were downloaded on March 11, 2022. It's important to note that the ranking list for each category changes by hours.

We manually navigated through each app, capturing 3 to 5 screenshots with different functionalities and layouts. In total, we collected 610 screenshots. However, given the high similarity in layout and icon design within the same categories, we discarded most of the screenshots, retaining 56 distinct screenshots representing diverse categories and unique icon types for our user studies.

\begin{figure}[h]
    \centering
    \includegraphics[width=0.4\textwidth]{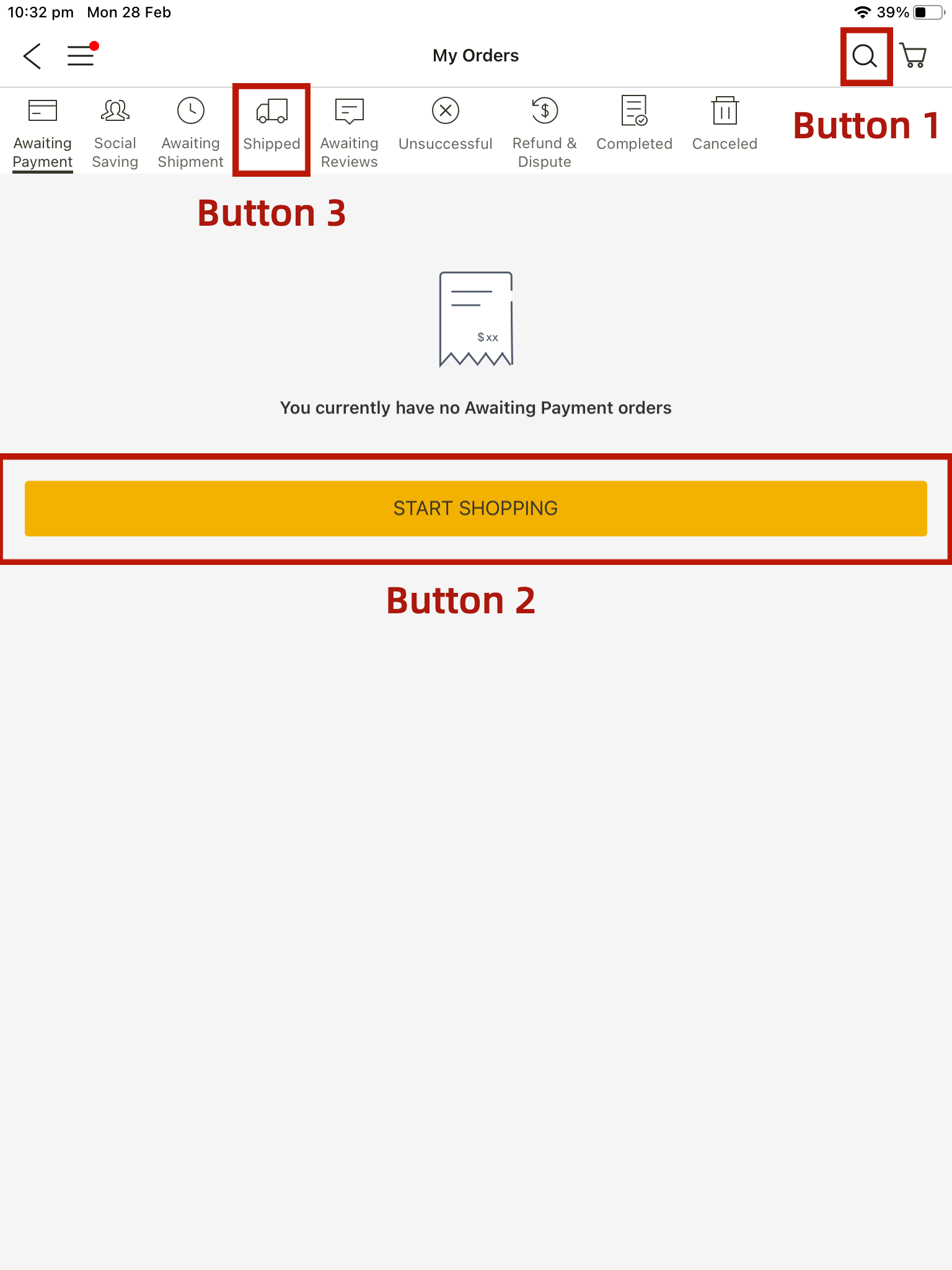}
    \caption{There are three types of buttons categorised by the contents of buttons. Button 1: with icon only; button 2: with text only; button 3: with both icon and text.}
    \label{fig_buttons}
\end{figure}

\begin{figure*}
    \centering
    \includegraphics[width=\textwidth]{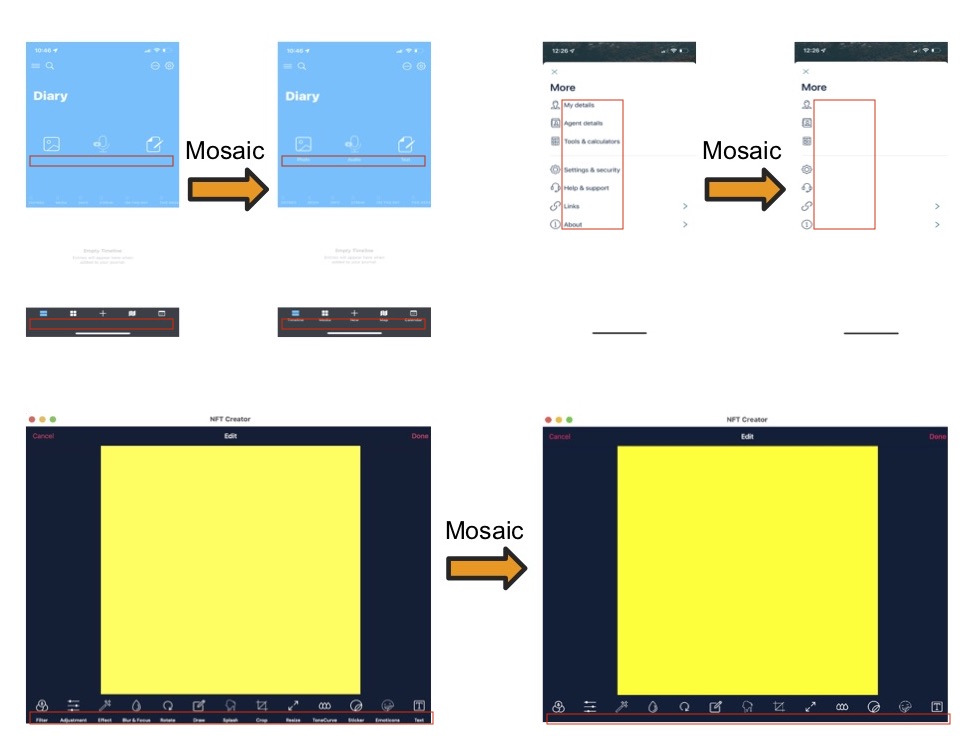}
    \caption{Illustrations of examples of mosaic application. After the mosaic processing, textual annotations of all icons on displayed UIs are cleared.}
    \label{fig_mosaic}
\end{figure*}

Buttons can be categorized based on their content into three types: icon buttons, text buttons, and icon+text buttons. For instance, 'Button 1' in Fig. \ref{fig_buttons} is an icon button, featuring an icon without any text describing its functionality. 'Button 2' is a text button, containing only text with no accompanying icon. 'Button 3' is an icon+text button, which can be activated by clicking either the text or the icon area.

However, during our UI exploration, we observed that for buttons with both text and icons, the text often did not correspond to the icon patterns. This discrepancy could potentially confuse users trying to discern the button's functionality based solely on the icon.

To further investigate this issue, we divided the 56 screens into two groups. Group 1, comprising 41 screens, predominantly featured icon+text buttons. We applied a mosaic effect to these screens, obscuring the text part of the icon+text buttons, leaving only icon buttons and text buttons visible. Fig.\ref{fig_mosaic} provides an example of this mosaic application.

Group 2 consisted of the remaining 15 screens, none of which contained icon+text buttons.

\begin{figure}
 \begin{subfigure}[t]{.5\textwidth}
    \centering
    \includegraphics[width=\linewidth]{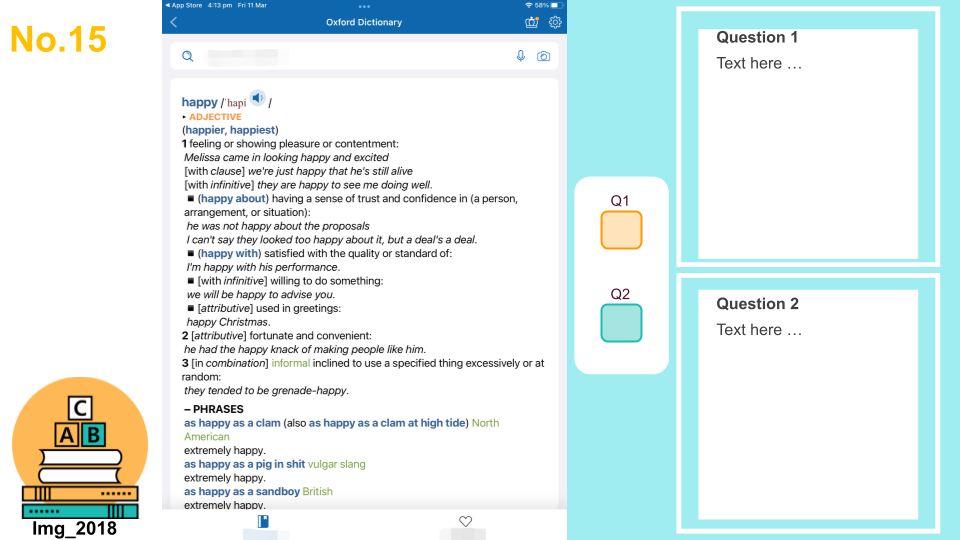}
    \subcaption{The first page for each UI, where users are required to select two icons on the displayed UI and drag the boxes with tags Q1 and Q2 to the corresponding icon area. Users then image the application scenarios of the displayed UI and propose an instance-level user intent for each selected icon.}
  \end{subfigure}
  \hfill
  \begin{subfigure}[t]{.5\textwidth}
    \centering
    \includegraphics[width=\linewidth]{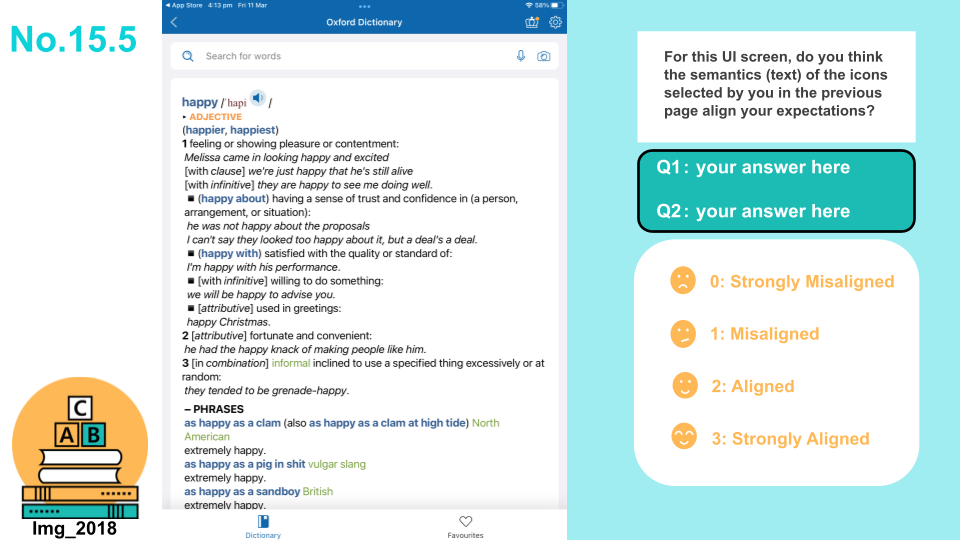}
    \subcaption{The second page for each UI, where users will see the textual annotations of icons which are hidden on the first page. Users are required to rate the alignment of the actual functionality of each selected icon with their expectations. }
  \end{subfigure}
  \vfill
    \begin{subfigure}[t]{.5\textwidth}
    \centering
    \includegraphics[width=\linewidth]{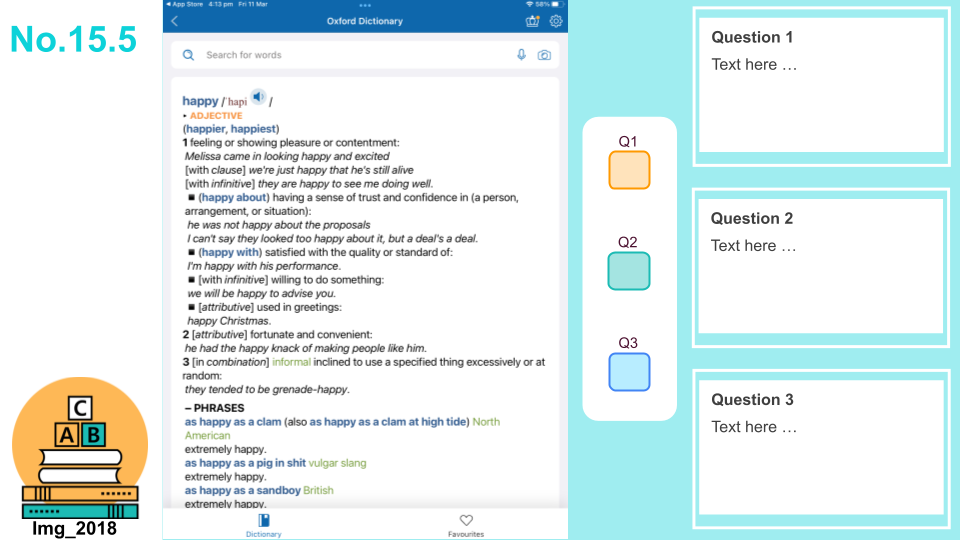}
    \subcaption{The third page for each UI, where users are required to propose instance-level intents for two icons in the same way as the first page, based on the textual annotations instead.}
  \end{subfigure}

  \caption{An example of the interface for the user study.}
  \label{fig_test_case}
  \vspace{-0.2in}
\end{figure}

\subsection{Procedure}

All user studies were conducted online, given that 50\% of our participants reside abroad, and due to the ongoing Covid-19 pandemic's hygiene policies. All participants consented to the publication of the collected dataset, including their names, contact information, age group, country of residence, spoken languages, and occupations.

Each user study began with a brief overview of our research tasks to ensure participants understood the study's goals. We then presented the study materials as online slides and explained the guidelines for providing answers on each slide. Excluding the pilot study, which contained 20 test cases, each participant's slide deck contained 17 test cases, each featuring a distinct screenshot from the selected screen dataset.

For Group 1 screens, where a subset of Mosaic screens was generated, the test case included three slide pages. The first page displayed the mosaic version of the screen, requiring users to guess the functionality of each icon (without text description) based on their experience. Users were asked to imagine a scenario where they interact with the displayed screen and consider their goal when clicking a specific button. After completing the first page, users viewed the original version of the Mosaic-modified screen on the second page. Here, they rated the alignment of each selected icon on a scale from 0 (Strongly Misaligned) to 3 (Strongly Aligned). On the third page, users proposed intents for selected icons on the original version of the screen, with text visible.

All user intents were expressed in English, with users encouraged to use various phrases such as \textit{`I wish to...'}, \textit{`I would like to...'}, \textit{`I'm going to...'}, etc. Fig.~\ref{fig_test_case} presents an example of a test case with the three types of pages mentioned above.

We conducted three versions of the user study, each involving three users. Before the official user studies, we conducted a pilot study with 20 test cases to gather feedback. Based on this feedback, we reduced the number of test cases in each user study to 17, allowing users more time to understand the UI content. We also adjusted the order of each test case to reduce overlapping icons between adjacent test cases, thus preventing users from proposing similar user intents without fully considering the potential functionalities of icons in a specific UI context. The remaining nine user studies were modified based on the feedback from the pilot study. 

\section{Evaluation}
\label{section_e}

\subsection{Testing Dataset}
\begin{figure}[h]
    \centering
    \includegraphics[width=0.45\textwidth]{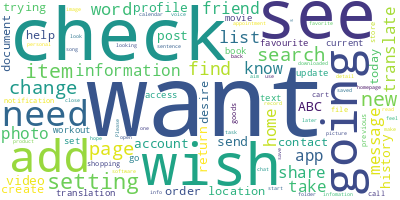}
    \caption{The Word Cloud of the top 100 words in the collected user intents in the testing dataset.}
    \label{fig_user_intent_cloud}
\end{figure}

The testing dataset collected from the user study comprises 752 distinct instance-level user intents, with $34.3 \%$ (258 out of 752) based on screens processed with the mosaic technique. Examples of these instance-level intents include: \textit{`This app is hard to use and I really need more instructions'}; \textit{`I would like to share my current location with my friends'}; \textit{`I would like to check my friends whose first name is Lucy'}; \textit{`I am really interested in this app and would like to know its background info'}; and \textit{`I am interested in fashion and would like to dive deep into that category'}, among others. The Word Cloud of these user intents is illustrated in Fig.~\ref{fig_user_intent_cloud}.

As previously mentioned, all textual elements were manually removed from mosaic screens, and the areas of the removed textual elements were filled with the background color of the text, as shown in Fig.~\ref{fig_mosaic}. This was done to evaluate our model's ability to identify target buttons on screens without annotations.

\begin{figure}[h]
    \centering
    \includegraphics[width=0.5\textwidth]{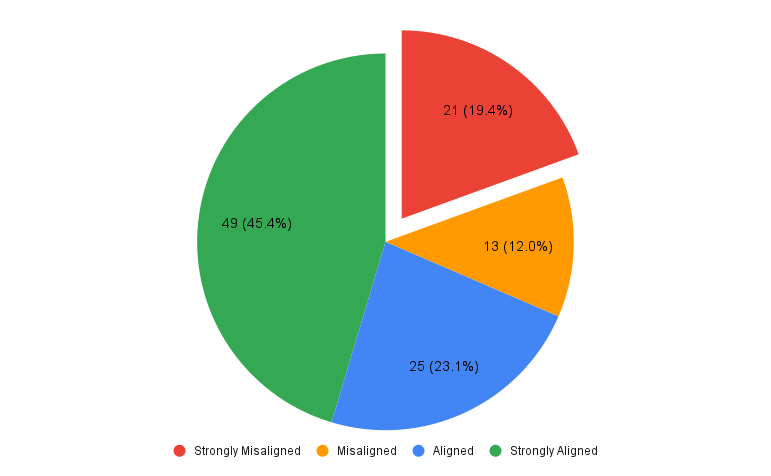}
    \caption{The alignment of actual functionalities of selected buttons on the mosaic screen with the user's expectations.}
    \label{fig_align}
\end{figure}

Among the user intents directed towards the mosaic screens, we also assessed the alignment between users' expectations and the actual functions as indicated by the annotations. As shown in Fig.~\ref{fig_align}, $19.4\%$ of target buttons were strongly misaligned with users' expectations, and $12\%$ were misaligned. Buttons that were aligned or strongly aligned with users' expectations accounted for $68.5\%$ of the total.

\subsection{Testing Automation}

\begin{figure}[h]
    \centering
    \includegraphics[width=0.2\textwidth]{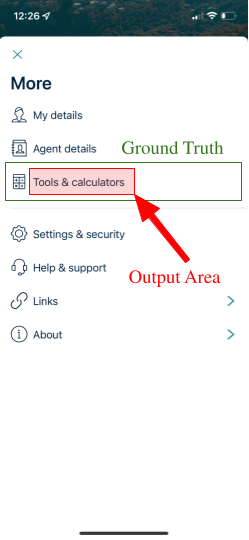}
    \caption{When the overlapping area of the output and the ground truth occupies more than 25\% of the output, the output is regarded as a correct result.}
    \label{fig_metrics}
\end{figure}

To automatically evaluate our model's performance, we generated a testing dataset from each user's provided intent. This was done by labeling the ground truth areas based on the areas framed by users in user studies. It's important to note that since a text+icon button can be activated by both the icon and the text area, we labeled the entire text+icon button as the ground truth if users framed only the icon or the text area of a certain button.

We also developed an evaluation agent to automatically compare our model's output with the ground truth of the testing dataset. The Intersection over Union (IoU) method, a widely used metric for measuring the accuracy of object detection tasks, is not applicable to our tasks. This is because the ground truth of our test cases includes both the text and icon area, while our model's output may contain only parts of that, i.e., the icon or text only. The icon itself or the text itself could be regarded as an accurate solution. Therefore, we proposed a metric where our model's output (target operational area) is considered accurate if the overlapping area of our output is no less than $25\%$ of the ground truth area. Fig.\ref{fig_metrics} provides an illustration, showing the ground truth of user content. The output of our model is also presented, which is verified as an accurate result since the overlapping area occupies more than $25\%$ of our output.

\subsection{Evaluating Capability of the Element Detector}

In our model, both the Icon Classifier and Descriptive-Word Classifier depend on the graphic or textual elements detected by UIED. Consequently, if UIED fails to detect the ground truth area, our model will miss it. Therefore, UIED's capacity to detect elements significantly influences our model's performance. However, UIED, equipped with customized GUI element detection methods, performs well with tailored configurations based on element scales on a specific screen, such as the size of the minimum GUI element. For a batch of different screens, finding a general configuration for UIED to optimize performance on each screen is challenging. For instance, a certain configuration of the minimum GUI element size may be too small, causing UIED to recognize partial button patterns or backgrounds as elements, or too large, leading UIED to miss many GUI elements on screens. The complexity of screen background also affects UIED's element detection accuracy.

\begin{figure}[h]
    \centering
    \includegraphics[width=0.5\textwidth]{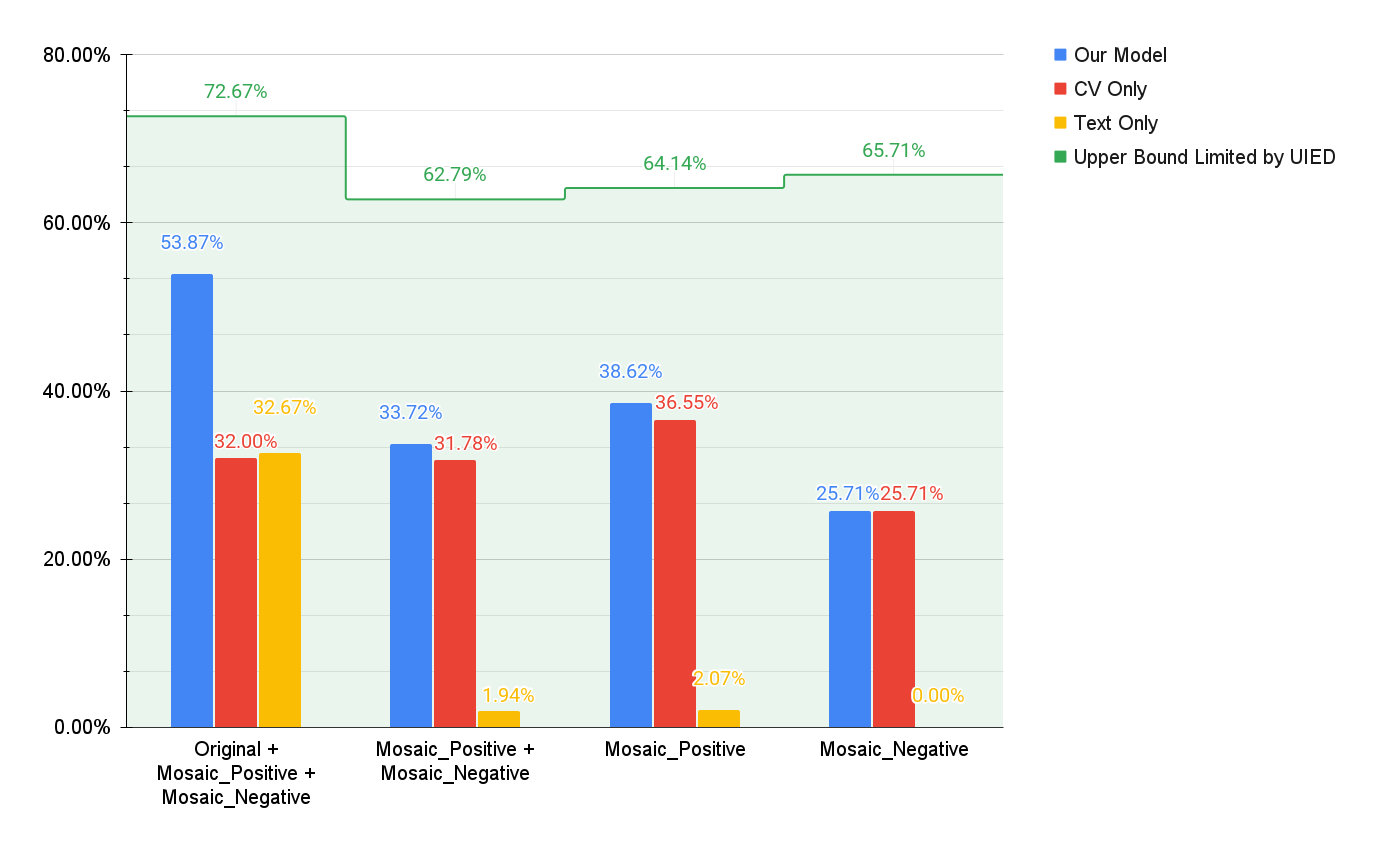}
    \caption{The accuracy of our model on datasets with mosaic screens.}
    \label{fig_eva_2}
\end{figure}

Utilizing our model's UIED configuration, we examined the proportion of testing datasets in which the ground truth area of each user intent is recognized as a standard GUI element by UIED on the corresponding screen. In the original screen dataset without the mosaic images, this proportion is $77.85\%$. This indicates that for $77.85\%$ of test cases, UIED correctly identifies their ground truth area as a standard GUI element. Conversely, for the remaining $22.15\%$ of test cases, UIED incorrectly classifies their ground truth area as background. Consequently, these missed ground truth areas are neither captured by UIED nor processed by our model in subsequent stages. In the Mosaic\_Positive dataset, the proportion of correctly identified ground truth areas by UIED is $74.73\%$. We further evaluated this proportion on other datasets: the dataset containing all user intents, the dataset encompassing all mosaic user intents, and the dataset comprising both positive and negative mosaic user intents. The respective proportions for these datasets are $72.67\%$, $62.79\%$, $64.14\%$, and $65.71\%$. These proportions are visually represented as the green area in Fig.\ref{fig_eva_1} and Fig.\ref{fig_eva_2}, corresponding to their respective testing dataset domains.

\subsection{Evaluating the End-to-End Operational Area Targeting}

\begin{figure}[h]
    \centering
    \includegraphics[width=0.5\textwidth]{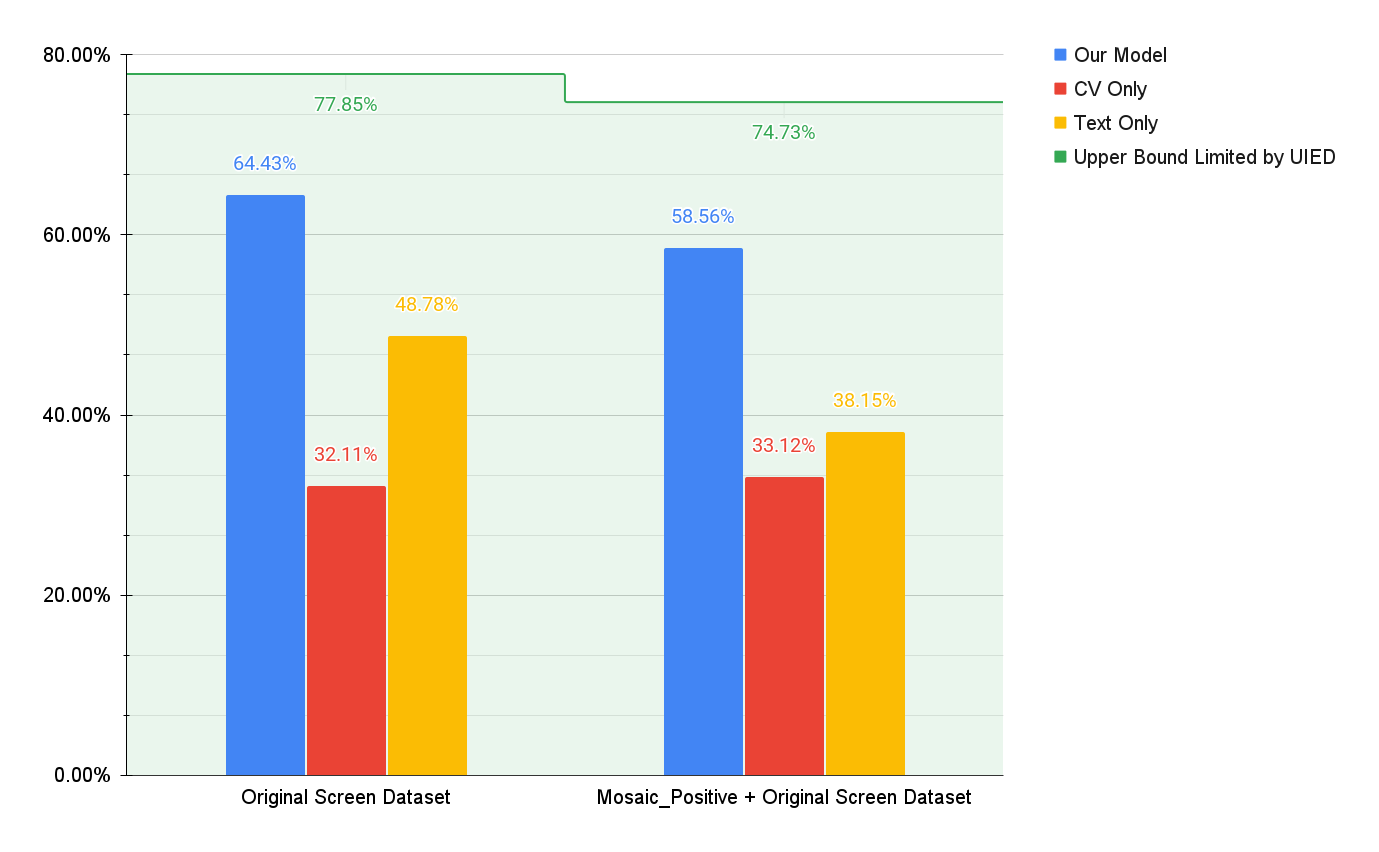}
    \caption{The accuracy of our model on the Original Screen Dataset and the Mosaic\_Positive + Original Screen Dataset}
    \label{fig_eva_1}
\end{figure}

\begin{figure}[h]
    \centering
    \includegraphics[width=0.5\textwidth]{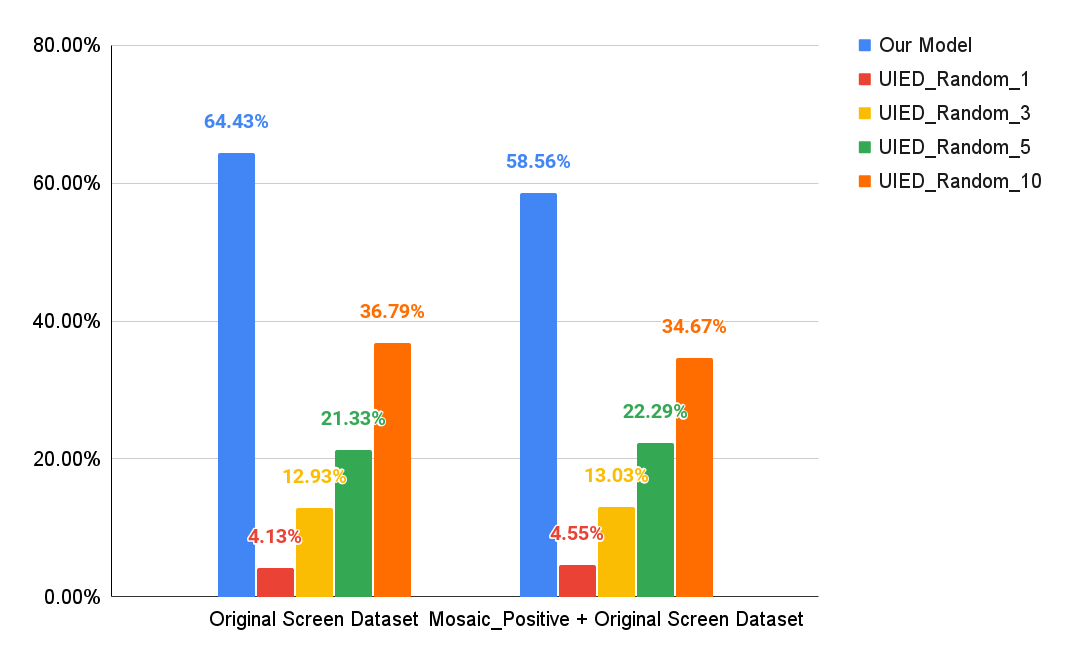}
    \caption{The accuracy of our model on the Original Screen Dataset and the Mosaic\_Positive + Original Screen Dataset compared with UIED\_Random Approaches.}
    \label{fig_eva_3}
\end{figure}
In our study, we categorized user intents in the testing dataset into two groups. The first group comprises user intents based on the original screens (without mosaic processing), which are collected when users proposed intents based on their understanding of the entire screen with annotation displayed. Consequently, the operational area recommended by users can be confidently regarded as the ground truth. Conversely, the second group includes user intents which are collected when the annotations were hidden from users. In this case, the recommended operational area by users is primarily based on their prior experience, without any reference to annotations. User intents rated as `misaligned' and `strongly misaligned' suggest that the recommended operational area cannot be treated as the ground truth. To evaluate the end-to-end operational area targeting ability of our model, we first tested our model on the original screens (referred to as \textbf{Original Screen Dataset}) and the dataset combining mosaic screens with 'strongly aligned' or 'aligned' feedback(referred to as \textbf{Mosaic\_Positive}), and the original screens. As shown in Fig.~\ref{fig_eva_1}, our model achieves an accuracy of $64.43\%$ on the \textit{Original Screen Dataset} and 58.56\% on the \textit{Mosaic\_Positive + Original Screen Dataset}. Note that, the notion `+' denotes the combination of multiple datasets. The decrease in accuracy with user intents on mosaic screens is due to the removal of all textual annotations, leaving only UIED and the computer vision (CV) techniques-based icon classifier to be adopted. We also compared the capability of our model with CV techniques only (ignoring textual elements detected by UIED) and the capability of our model with Text only (ignoring graphic elements detected by UIED). As depicted in Fig.~\ref{fig_eva_1}, when textual contents are ignored, the accuracy of the model on the original dataset is $32.11\%$, and the accuracy on the Mosaic\_Positive + Original Screen Dataset increases to $33.12\%$. If only textual elements are considered while ignoring the graphic elements on screens, the accuracy on the original screen dataset is $48.78\%$ and drops to $38.15\%$ on the Mosaic\_Positive + Original Screen Dataset.

\subsection{Evaluating Capability on Text-Free Screens}

We further evaluated our model's capability on the screens without textual annotations. 

As textual elements cannot be searched, the accuracy of our model on the three mosaic datasets decreases significantly compared to the accuracy on the testing dataset with the original screen + Mosaic\_Positive dataset. The inclusion of negative mosaic data reflects practical scenarios where many buttons' functionalities are not aligned with users' expectations (referred to as \textbf{Mosaic\_Negative}). As shown in Fig~\ref{fig_eva_2}, with the negative mosaic data, our model's accuracy is $53.87\%$, where the CV technique only can accomplish $32\%$ of accuracy, and the Text technique only can achieve $32.67\%$. For the dataset where all annotations of graphic elements are removed, our model achieves $33.72\%$ on the Mosaic\_Positive+Mosaic\_Negative, $38.62\%$ on the Mosaic\_Positive, and $25.71\%$ on Mosaic\_Positive, respectively.

\subsection{Usability Assessment}

To assess our model's usability, we compared its performance with a random algorithm that randomly selects a certain number of elements from elements detected by UIED. This random algorithm simulates a situation where a layman does not know the targeting operational area and randomly explores buttons on the screen. We denote UIED\_Random\_x as the maximum number of randomly selected elements of the approach to cover the targeting area. Fig.~\ref{fig_eva_3} shows the accuracy of each approach, indicating the possibility of covering the targeting operational area when randomly selecting a certain number of elements. The results illustrate that our model significantly outperforms the random approaches.

\section{Discussion}

The GUI element detector is crucial in providing target element candidates for user intents, as demonstrated in Section~\ref{section_e}. The performance of candidate detection significantly influences our model's accuracy on the testing dataset. To the best of our knowledge, no existing GUI detection tools are designed for a range of general UIs with different scales. This reliance on the capabilities of existing tools is a limitation not only for our approach but also for all pipeline-based approaches. However, theoretically, our model's performance could improve with a more capable GUI detection tool.

Additionally, many icons serve merely as decorations and do not indicate specific functionalities. These icons often differ significantly from user expectations, as indicated in fig.~\ref{fig_align}, limiting our model's accuracy on the testing dataset, especially for UIs without any textual annotation. This limitation could be mitigated by incorporating more icons with customized styles into the CNN icon classifier of the pipeline.

Our approach has various potential applications. For instance, it could make smartphone services more accessible to individuals unfamiliar with smartphones, those who find them inconvenient to use, or those with no smartphone knowledge, such as older adults and people with disabilities. Users can easily interact with smart devices by describing their situation, which motivates our approach to process instance-level user intents. Our approach could also save users time when exploring new apps by quickly identifying the target area within two seconds.

From a machine learning perspective, our approach represents a significant breakthrough by bridging linguistics and computer vision, translating instance-level user intents into precise operating areas on UIs. The model presented in this paper could be directly adopted for generating training data for the convergence of Computer Vision (CV) and Natural Language Processing (NLP).

\section{Conclusion}

In this paper, we have explored the capability augmenting of virtual assistants by accurately targeting operational areas on user interfaces (UIs) to address instance-level user intents. We proposed a novel cross-modal pipeline that combines natural language processing and computer vision techniques through a unique cross-modal bridging approach. Furthermore, we collected a dataset featuring instance-level user intents and precise operational areas on UIs from a diverse group of mobile users for automation testing. Our model achieved an accuracy of 64.43\% on this testing dataset, demonstrating its ability to accurately and comprehensively understand instance-level user intents and identify the precise operational area without relying on metadata. This result underscores our approach's success in achieving its goal: requiring zero effort from developers and end-users, handling custom and instance-level tasks, and demonstrating adaptability across any app category.

\bibliographystyle{ACM-Reference-Format}
\bibliography{uist}

\end{document}